# Deep Barcodes for Fast Retrieval of Histopathology Scans

Meghana Dinesh Kumar[1], Morteza Babaie[2] and H.R. Tizhoosh[1]

[1] KIMIA Lab, University of Waterloo, Ontario, Canada
[2] Mathematics and Computer Science, Amirkabir University of Technology, Tehran, Iran

**ABSTRACT**

**We investigate the concept of *deep barcodes* and propose two methods to generate them in order to expedite the process of classification and retrieval of histopathology images. Since binary search is computationally less expensive, in terms of both speed and storage, deep barcodes could be useful when dealing with big data retrieval. Our experiments use the dataset *Kimia Path24* to test three pre-trained networks for image retrieval. The dataset consists of 27,055 training images in 24 different classes with large variability, and 1,325 test images for testing. Apart from the high-speed and efficiency, results show a surprising retrieval accuracy of 71.62% for *deep barcodes*, as compared to 68.91% for *deep features* and 68.53% for *compressed deep features*.**

## 1. INTRODUCTION

*Digital pathology* deals with high-resolution scans (images) of biopsy specimen. Managing and using information generated from digital pathology scans could be done through computerized methods. Due to the recent introduction of Whole-Slide Imaging (WSI), digital pathology has become one of the most promising fields of diagnostic medicine. Digitized tissue pathology is open to numerous applications of machine learning and image analysis techniques. When physicians are examining medical images, retrieving similar images from a large archive could provide useful information to support a diagnosis. However, since archives of medical images, particularly in digital pathology, contain a prohibitively large number of gigapixel files, conventional search methods will most certainly fail (a typical resolution is one pixel per 0.5 micron resulting in most scans to contain gigabytes of information). It will be useful if each image, or its regions/patches, is tagged with a "barcode" for a fast retrieval. The storage requirement will also reduce, enabling more data for indexing and reference. In this paper, we propose novel compact descriptors for medical image retrieval. We used the largest publically available dataset of pathology images called *Kimia Path24*. It contains 27,055 images for training/indexing and 1,325 images for testing [35]. Different experiments were performed on three pre-trained networks on non-medical images (ImageNet): AlexNet [30], VGG-16 and VGG-19 [31], [32].

Section 2 provides a brief review of relevant literature, among others works of neural networks and binary descriptors. Section 3 introduces deep barcodes. Experiments are described in Section 4. Section 5 reports the results which are subsequently discussed in Section 6. The paper is concluded in Section 7.

## 2. BACKGROUND

Computer algorithms are commonly used in medical imaging to complement assist physicians in their decision-making [1,2]. Pathology slides provide a more complete view of diseases and effects on tissues as the process by which slides are prepared preserve the essential tissue architecture [3]. Moreover, a few disease characteristics like lymphocytic infiltration of cancer can be inferred only by pathology images. Furthermore, analysis of pathology images continues to be considered as 'gold standard' in diagnosing many diseases which involve all types of cancer [4]. The exploration of spatial structure in pathology images can be tracked back to the studies by Wiend *et al*. [5] in 1998, Bartels *et al*. [6] in 1992 and Hamilton *et al*. [7] in 1994. This domain was mainly overlooked due to the scarcity of computational resources and expensive equipment for digital imaging. However, assessment of spatial features in pathological images has recently become the backbone of most automated pathology image analysis techniques.

Gurcan *et al*. [3] have presented a comprehensive review of various state-of-the-art CAD methods and practices that digitized pathology images employ for automatic image analysis. Ciresan *et al*. [37] developed a powerful pixel classifier using DNN, which is a max-pooling convolutional neural network. The input to their DNN is only raw RGB data, hence there is no need for human input, which makes the DNN automatically learn visual features from training data.

Input is sampled from a square patch of source image, centered on the pixel itself. The network is trained to distinguish patches that contain mitotic nucleus near the middle of the window. Petushi *et al.* [9] experimented with a scheme for tissue micro-texture classification of breast cancer by extracting textural features and segmenting nuclei. The two textural features were: the surface density of nuclei and spatial position. Naik *et al.* [10] showed a technique which detects and segments glands and nuclei in breast cancer pathology images. They used 51 graph-based features from minimum spanning tree, Voronoi diagrams and Delaunay triangulation. Their method employed nuclear centroids to discriminate different regions of cancer from benign areas on whole slide images, yielding an overall accuracy of 80% in conjunction with a support vector machine classifier. Cruz-Roa *et al.* [8] used a DL framework that extends many convolutional neural networks for visual semantic analysis of regions prone to tumor for diagnosis support. Their CNN is trained over a huge number of whole slide scan patches (different tissues), thereby learning a hierarchical part-based representation.

For several years, microscopic information of specimens has been archived and stored using glass slides [23]. These glass slides are very fragile and moreover, they require particularly equipped storage rooms which occupy large area, to store specimens, leading to high logistical infrastructures. Simplified representations along with techniques to learn features that need not incorporate domain knowledge have been used in complex learning tasks [17, 19]. These techniques completely prevent segmentation and selection of handcrafted features manually, thereby addressing learning tasks like *learn-from-data* approach [18]. Zhang et al. [24] introduced a scalable CBIR method to deal with WSI using a supervised kernel hashing technique. This hashing technique reduces a 10,000-dimensional feature vector to 10 binary bits. Even after the decrease in size of features, the method was seen to preserve a succinct representation of the image. These abridged binary codes are used to index all existing images in the dataset for quick retrieval of new query images. The efficiency of their proposed algorithm is demonstrated on a breast pathology dataset, which is obtained from 116 patients, and comprised of 3,121 WSI scans.

## 2.1. Deep Neural networks

Lately, deep learning architectures have become very popular, owing to their success in many computer vision and pattern recognition tasks. They are known to outperform conventional methods including handcrafted features for data representation and machine learning tasks [11, 12, 13, 14, 15, 16]. One of the main reasons for this is the exponential growth of digital data available, along with powerful computational resources [20]. Essentially, they can be considered as evolved version of conventional neural networks [17, 18]. CNNs are a part of multilayer neural networks, which are specially designed to be used on two-dimensional data like images and video frames [18]. Since they are feed-forward networks, their architecture is built in such a way so as to decrease sensitivity to distortions in input image, such as translations and rotations.

It is well understood that deep learning approaches are especially proficient at predicting outcomes, especially if a large number of training samples are available. Additionally, this would also ensure the generalizability of learned features and classifiers. The learning approach involves multiple operations on data, including non-linear transformations, which yields abstract and handy representations [19]. In 2015, Bar *et al.* [29] studied the application of deep learning approaches for pathology detection in chest radiograph data. They also explore the possibility of using convolutional neural networks (CNNs) that are learned from ImageNet, which is a large scale non-medical database. The experiments demonstrate that best performance is achieved by combining both deep learning (Decaf) and PiCodes ("Picture Codes").

One of the major advantages of using deep learning is that while training, models learn highly suitable representations in hierarchical manner. This is analogous to the way pathologists analyze pathology slides over varying resolutions and fields of view. Nonetheless, our deep learning strategy, like moth others, employs deep learning based over small parts of the image. This way, the classifier network is applied on all parts of the whole slide scan canvas. Our data consists of images from *Kimia Path24 dataset*, which is useful for studying image classification and retrieval in digital pathology. It consists of 1325 images for testing and 27,055 images for testing. Since the Deep Neural Network (DNN) operates on raw pixel values, there is no necessity for human input; the DNN automatically learns a set of visual features from the training data. Classification and retrieval of images in medicine are very critical tasks. They are used to assist medical practitioner in making informed decisions regarding patient's medical case. Digitizing pathology images provides many advantages. They do not decay over time and can be examined by several specialists throughout the world at the same time. Additionally, they can be retrieved very easily if the image processing algorithms are efficient. This properly can be exploited by researchers and quality control personnel.

Digital pathology, which is digitized form of pathology glass slides, is among the latest instances of big electronic data [21]. Each of these images, i.e., whole scans of pathology samples are several gigabytes [22]. Therefore, they are difficult to store, process and transfer whole images in real time. Learning deep features belonging to such large digital pathology images is a good chance to learn hidden patterns which cannot be observed by humans.

## 2.2. Pre-trained networks

We used three major pre-trained networks: AlexNet, VGG-16 and VGG-19. AlexNet is a Convolutional Neural Network (CNN) that is trained on approximately 1.2 million images belonging to ImageNet Dataset [33]. The architecture of this network 23 layers and it can classify images into 1000 categories. Similarly, VGG-16 and VGG-19 have been trained on a subset of images belonging to the ImageNet database, which are used in ImageNet Large-Scale Visual Recognition Challenge (ILSVRC) [32]. These two networks are trained over one million images, and can classify input images into 1000 different categories. Due to the training process, all the three models have learned rich feature representations for a vast range of input image.

Since pre-trained networks are trained on other domains and classify input images into different categories which are irrelevant for our current task, we use distance metrics as classifiers based on best similarity (least distance). Deep Neural Networks (DNN) operates on raw pixel values, there is no necessity for human input; the DNN automatically learns a set of visual features from the training data.

**2.3. Binary descriptors**

Several studies show that binary features can be extremely powerful and concise, and are able to compete with the state-of-the-art complex learning algorithms. The most relevant study performed by Tizhoosh [38] showed how barcodes could facilitate image retrieval. It proposed generation and use of barcodes for annotation in the medical imaging domain. The study investigated different regions of interest such as organs, tumors and tissue types. It introduced Radon barcodes (RBC), and implemented local binary patterns (LBP) and local Radon binary patterns (LRBP) as binary barcodes. Validation was carried out on IRMA x-ray dataset.

## 3. Deep Barcodes

Deep features have shown promise for recognition of objects and faces in many applications. However, as pathology scans are of large dimensions, they need to be split into a large number of smaller images to become manageable. A scan can easily produce several hundred to several thousand sub-images (patches) that are still large (e.g., 1000x1000 pixels). A feasible image retrieval system cannot work on real-valued feature vectors for the scan. Hence, we binarize *deep features* to obtain *deep barcodes*. Each deep barcode has 4096 features in case VGG network. Binary search is extremely fast and requires less resources, which is the primary motivation behind this paper.

**Stage 1** - We compute a *deep barcode* for each image. The barcode is calculated in two ways:

(a) *Min-max algorithm*
In this case, deep barcode output *f(x)* is 1 if there is an increase in the value of consecutive feature (*x*), and 0 otherwise. This way, *f(x)* is of dimension 1x4095, where *n* belongs to [1,4095]. The min-max algorithm can be represented as:

$$f(x_n) = \begin{cases} 1, & x_n < x_n + 1 \\ 0, & x_n \geq x_n + 1 \end{cases} \quad (4)$$

This process is repeated on all images in database, including the training and testing sets;

(b) *Zero-thresholding*
After obtaining the *deep features*, we threshold the entire feature set; whatever lies above zero is assigned 1 and the feature values below or equal to zero are assigned 0. This way, each image has a feature of dimension 1x4096.

$$g(x_n) = \begin{cases} 1, & 0 < x_n \\ 0, & 0 \geq x_n \end{cases} \quad (5)$$

Again, this process is repeated for all images in the training and testing data sets. Therefore, each image has a barcode of 4096 digits. Fig. 1 shows two sample images with their deep features and deep barcodes.

Upon obtaining the barcodes, we compute the distance of each barcode in the test set to each barcode in the training set. This distance is computed using logical XOR metric, by comparing the number of differences between corresponding elements between two barcodes. Using the method mentioned above, we obtain the best match of each test barcode with a barcode in training set. This is the *first stage search*; which yields top *N* matches.

**Stage 2** - The input to second stage search is the test images along with their top *N* matched images (using their barcodes and XOR distance comparison) from training set. The purpose of stage 2 search is to find only *one* best match, out of the top *N* matches. *Deep features* obtained by extracting activations from deepest layer in the trained network is compared in this stage. We perform this using two distance metrics: city-block distance ($L_1$ norm) and Euclidean distance ($L_2$ norm). They help in (dis)similarity measurement when two feature vectors are compared. Hence, we obtain a single best match (from training set) for each and every 1325 images in test set.

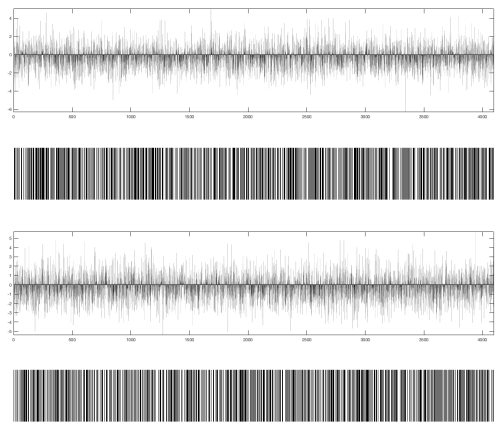

**Fig. 1**. Visualization of deep features (top) and corresponding barcodes (bottom) for two sample images.

## 4. EXPERIMENTS

Majority of the previously proposed methods are usually based on handcrafted visual features (like color and texture) which are used to simulate visual perception of human pathologists to understand tissue samples [26]. Deep Learning has been recently applied in this field, to classify and analyze big data. Babaie et al. [34] suggested that both feature extractor and transfer-learned networks were able to offer an increase in classification and retrieval accuracy on the *Kimia Path24* dataset when compared to a CNN trained from scratch. We used three such well-known networks, which are pre-trained on non-medical images: AlexNet [30], VGG-16 and VGG-19 [31], [32].

The 27,055 images were fed into each of the three networks separately. From the deepest layer, features were extracted and these are referred to as *deep features*. Each feature is a vector of dimension 1x4096. Therefore, the deep features for training dataset have a dimension of 27055x4096, while the deep features of testing dataset have a dimension of 1325x4096.

### 4.1. Dataset

*Kimia Path24* is a uniquely designed dataset, used for classification and retrieval of digitized pathology images. A few images from this dataset are shown in Fig. 2. The dataset exhibits a large variability in patterns. Initially, there were 350 whole scan images (WSIs) from diverse body parts for use. All of these whole scan images were captured by *TissueScope LE 1.0* [36]. The images were scanned using a 0.75 NA lens in the bright field. Each scan has a header that contains image information like resolution. Out of 350, only 24 WSI scans are selected that consist of different tissue textures based on visual examination for non-medical practitioner. Moreover, it is a conscious decision to select a subset of WSI such that each of them characterize different textural patterns (different tissue types) as most public datasets for histopathology focus on cancer classification via yes/no decisions.

Each of the WSI scans is divided into multiple patches, each of size 1000x1000 pixels. Among the possible patches, 1,325 number of patches are manually selected, with particular consideration to textural differences (high variability). This way, the users of this dataset have the option to create many training patches, ranging from 27,000 to more than 50,000 patches; we used 27,055 patches for training. The number of training patches depend on the selection of homogeneity and overlap parameters for each slide. For retrieval performance, a weighed accuracy measure is specified to enable unified benchmark for future work. *Kimia Path24* is the largest dataset available to our best knowledge, and is released to the public. We have performed thorough experiments on this dataset to demonstrate the efficiency of our proposed approach. The *Kimia Path24* dataset can be downloaded from the website of KIMIA Lab [35].

#### 4.1.1 Accuracy Calculation

We validate our algorithm by computing its performance of classification and retrieval, using three different *accuracy* measures. Each test image and its closest match should belong to the same category, i.e., among the 24 categories of *Kimia Path24* dataset. There are a total of $n_{tot} = 1,325$ patches $P_s^j$ that belong to 24 sets $\Gamma_s = \{P_s^i | s \in S, i = 1,2...,n_{\Gamma_s}\}$ with $s = 0, 1, 2, \ldots, 23$. Looking at the set of retrieved images $R$ for any experiment, the patch-to-scan accuracy $\eta_p$ can be given as:

$$\eta_p = \frac{\sum_{s \in S} |R \cap \Gamma s|}{n_{tot}} \quad (1)$$

We also calculate the whole-scan accuracy $\eta_W$ as:

$$\eta_W = \frac{1}{24} \sum_{s \in S} \frac{|R \cap \Gamma s|}{n_{\Gamma_s}} \quad (2)$$

The total accuracy $\eta_{total}$ takes into account both patch-to-scan and whole-scan accuracies:

$$\eta_{total} = \eta_p \times \eta_W \quad (3)$$

The Matlab and Python code for accuracy calculations can also be downloaded [34], [35].

### 4.2. Experiment 1: Deep features

Initially, we used the deep features obtained from the activations of deepest layer in each network. Similarity between a pair of features (one feature taken from the test set and the other from the train set) was compared using city-block distance ($L_1$ norm) and Euclidean distance ($L_2$ norm). For each feature in test set, its closest match (least distance) from training set was found. Using this concept, we computed accuracy of classification and retrieval of all 1,325 images.

### 4.3. Experiment 2: Principal component analysis

In order to reduce the dimensionality, and obtain compact features, we applied principal component analysis (PCA). We computed the principal components of training images and used them to represent images in the test dataset. PCA was applied individually on deep features of all three pre-trained networks, VGG-16, VGG-19 and AlexNet. The number of principal components used for retrieval were varied, between 20 to 200 components. Additionally, we also studied its variation with two types of distance metrics: $L_1$ norm and $L_2$ norm. These metrics were used to compare the similarity between two features upon reducing their dimensionality. We address this as *reduced deep features*.

### 4.4. Experiment 3: PCA with Deep Barcodes

The third experiment involved multi-stage search using principal component analysis as well as deep barcodes. In the first stage, we computed *reduced deep features*. These features were binarized using the *min-max* technique and *reduced deep barcodes* were computed. Therefore, not only were the dimensions of features reduced, but they were also binarized. The output of first stage was a set of $N$ images whose *reduced deep barcodes* are most similar to the test image. In the second stage search, we used $L_1$ distance metric to compare deep features among the top $N$ matches, yielding a single best match image from training set. We conducted this experiment using VGG-16 pre-trained network and $L_1$ distance metric for the second stage search.

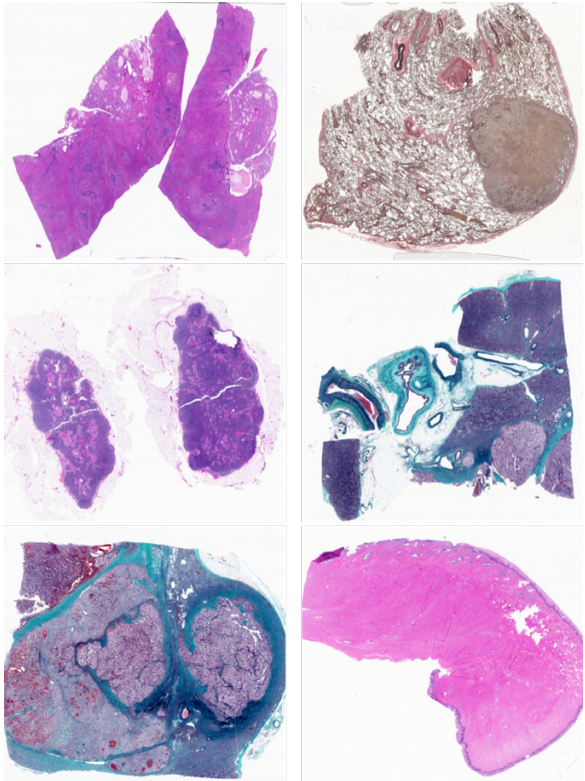

**Fig. 2:** Sample images from the *Kimia Path24* dataset. Aspect ratios have been neglected for better illustration [34].

### 5. RESULTS

In the first experiment, the deep features of test set were compared with all deep features of training set using $L_1$ and $L_2$ distance metrics. In this experiment, we used the conventional procedure, without PCA and no barcodes. The results are shown in Table 1.

In the second experiment using PCA, we achieved the highest accuracy ($\eta_p$) of 68.53%, which is comparable with the results obtained without using any processing on deep features. Results are shown in Table 2.

We used binary search for comparing deep barcodes and obtained comparable classification and retrieval accuracies, and even higher (than deep features and reduced deep features) in certain cases. The value of $N$ was varied between 1 and 1000. Note that $N$ is the number of matches output from stage 1 search. Barcodes were obtained using the min-max algorithm. The results for different architectures are shown in Table 3.

Since VGG-16 provided the best classification and retrieval accuracies, we continued to study it in detail. Table 4 shows these results. Tables 5 and 6 show the results of experiment 4, i.e., retrieval accuracy using *reduced deep barcodes*.

In tables 2-6, $N_{PCA}$ denotes the number of principle components used, along with the distance metric for similarity measure. $N$ is the number of top matches obtained in the first stage search, which are fed into the second stage search, to find a single best match.

| Network | Metric | $\eta_p$ | $\eta_W$ | $\eta_{total}$ |
|---|---|---|---|---|
| AlexNet | L1 | 61.89 | 59.44 | 36.78 |
|  | L2 | 61.96 | 59.23 | 36.70 |
| VGG-19 | L1 | 67.32 | 65.38 | 44.01 |
|  | L2 | 67.70 | 65.99 | 44.68 |
| VGG-16 | L1 | 68.23 | 66.92 | 45.66 |
|  | L2 | 68.91 | 67.50 | 46.51 |

*Table 1: Retrieval accuracies ($\eta_p$) using deep features from pre-trained networks (without dimensionality reduction, no binarized features).*

### 6. DISCUSSIONS

We observe that the best performance was achieved by VGG-16 pre-trained network architecture, in the case of deep features, reduced Deep features and deep barcodes. Additionally, this experiment is a novel approach demonstrating that deep learning trained with large scale non-medical image databases may provide good recognition capability even in the case of pathology images.

It is astonishing to see that the classification and retrieval accuracies obtained using *reduced deep features* and *deep barcodes* were close to the accuracies of search and retrieval using non-compact *deep features* (and even better in certain cases). Additionally, when we select a *single* top match (single stage search) comparing only deep barcodes among the entire 27,000 images, we achieved the highest retrieval accuracy. It is also important to note that we simply used the architectures from pre-trained networks (which were trained on a non-medical image dataset, namely the *ImageNet*). This demonstrates that there is minimum (or *no*) loss in information when the activations of deepest layer are converted to *deep barcodes* with only binary values. Hence, further computation and developments can possibly exploit this fact to tackle many challenges, especially the challenge of resource requirements. There is increased speed since only binary values have to be compared.

For a clear understanding, Fig. 3 graphically represents the variation of accuracy ($\eta_p$) by applying min-max binarization technique. Fig. 4 graphically represents the results of the experiment, i.e., variation of total accuracy $\eta_{total}$ when different number of PCA components are considered in the first stage search ($N_{PCA}$).

## 6.1. Comparison with existing methods

We compare our results with the previous work on the same dataset carried out by Babaie et al. [34]. The authors have used three different methods for retrieval, Local Binary Patterns (LBP), Bag-of-words (BOW) and two CNNs that are trained from scratch. Our results are better than all the three methods they have reported, with an additional advantage of reduced dimensionality and computational expense. Table 6 summarizes these results.

## 7. CONCLUSION

The aim of this entire proposal may probably be considered very relaxed, since we are simply trying to classify the patches into the respective whole scan images. However, as clearly seen in the reported experiments and previous work, this task is extremely challenging. We examined whether deep nets with ImageNet training may be suitable for general medical image recognition tasks. We produced compact features for image retrieval, *reduced deep features* and *deep barcodes*, both derived from *deep features* of a pre-trained network. Moreover, deep barcodes which are feature vectors with binary values, and can easily be used for retrieval of images using binary search techniques. The retrieval accuracy reached 71.62% which is higher than that obtained using the real-valued deep features.

deep networks trained from scratch. However, a possible reason behind this may be the number of images the deep network was trained on. 27,055 images seems to be a very small sample to train a deep network. However, due to the lack of labelled data and a larger dataset on pathology images, *Kimia Path24* seems to be the best dataset to validate our algorithm.

There are certain difficulties faced in using deep learning for pathological images, among others a lack of large labelled image databases, decreasing the resolution of patches so that it may be used with deep network, and the fact that the pre-trained networks we used were originally trained on non-medical images; if they were trained on medical images, we could have possibly achieved higher retrieval accuracy since the architecture would be well adjusted for our purpose.

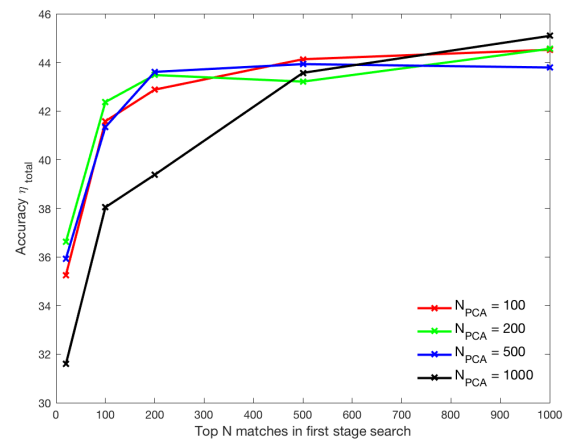

**Fig. 4**. variation of total accuracy $\eta_{total}$ when different number of pricipial components are considered.

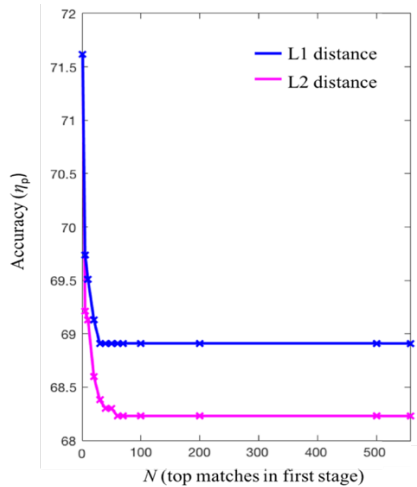

**Fig. 3.** Variation of retrieval accuracies for N matches, i.e., the number of top matches in first stage search. Deep barcodes are generated using the min-max algorithm.

Babaie et al. [34] presented that networks that are pre-trained on non-medical images perform better compared to

| Network | $N_{PCA}$ | Metric | Accuracy | | |
|---|---|---|---|---|---|
| | | | $\eta_p$ | $\eta_W$ | $\eta_{total}$ |
| VGG-16 | 20 | L1 | 61.28 | 59.59 | 36.32 |
| | | L2 | 61.81 | 59.27 | 36.63 |
| | 50 | L1 | 68.53 | 67.16 | 46.03 |
| | | L2 | 68.53 | 67.13 | 46.00 |
| | 200 | L1 | 65.81 | 64.71 | 42.58 |
| | | L2 | 68.53 | 67.15 | 46.01 |
| VGG-19 | 20 | L1 | 62.26 | 59.77 | 37.21 |
| | | L2 | 62.11 | 60.00 | 37.26 |
| | 50 | L1 | 67.25 | 65.03 | 43.73 |
| | | L2 | 66.87 | 65.07 | 43.51 |
| | 200 | L1 | 65.43 | 64.08 | 41.93 |
| | | L2 | 67.40 | 65.73 | 44.30 |
| AlexNet | 20 | L1 | 57.81 | 53.90 | 31.16 |
| | | L2 | 57.96 | 54.75 | 31.73 |
| | 50 | L1 | 61.13 | 58.01 | 35.46 |
| | | L2 | 62.34 | 59.20 | 36.90 |
| | 200 | L1 | 58.57 | 56.26 | 32.95 |
| | | L2 | 62.64 | 59.89 | 37.52 |

Table 2: Retrieval accuracies ($\eta_p$) using reduced deep features by applying PCA.

| Network | $N$ | Metric | $\eta_p$ |
|---|---|---|---|
| AlexNet | 20 | L1 | 62.26 |
| | | L2 | 61.96 |
| | 100 | L1 | 61.89 |
| | | L2 | 61.96 |
| VGG-19 | 20 | L1 | 67.55 |
| | | L2 | 67.77 |
| | 100 | L1 | 67.32 |
| | | L2 | 67.70 |
| VGG-16 | 20 | L1 | 68.60 |
| | | L2 | 69.13 |
| | 100 | L1 | 68.23 |
| | | L2 | 68.91 |

Table 3: Retrieval accuracies ($\eta_p$) obtained using two-stage search with min-max binarization.

| $N$ | Min-max | | Zero-threshold | |
|---|---|---|---|---|
| | L1 | L2 | L1 | L2 |
| 1 | 71.62 | 71.62 | 70.64 | 70.64 |
| 5 | 69.21 | 69.74 | 69.43 | 69.89 |
| 10 | 69.13 | 69.51 | 68.91 | 69.66 |
| 20 | 68.60 | 69.13 | 68.38 | 68.98 |
| 30 | 68.38 | 68.91 | 68.30 | 68.98 |
| 40 | 68.30 | 68.91 | 68.23 | 68.91 |
| 50 | 68.30 | 68.91 | 68.23 | 68.91 |
| 60 | 68.23 | 68.91 | 68.23 | 68.91 |
| 70 | 68.23 | 68.91 | 68.23 | 68.91 |
| 100 | 68.23 | 68.91 | 68.23 | 68.91 |
| 500 | 68.23 | 68.91 | 68.23 | 68.91 |

Table 4: Retrieval accuracies ($\eta_p$) obtained using deep barcodes by min-max and zero-threshold algorithms with VGG-16.

| $N$ | $N_{PCA}$ values | | | |
|---|---|---|---|---|
| | 100 | 200 | 500 | 1000 |
| 20 | 35.25 | 36.62 | 35.93 | 31.59 |
| 100 | 41.58 | 42.37 | 41.34 | 38.04 |
| 200 | 42.88 | 43.48 | 43.61 | 39.38 |
| 500 | 44.13 | 43.21 | 43.93 | 43.57 |
| 1000 | 44.52 | 44.56 | 43.79 | 45.09 |

Table 5: Total accuracy ($\eta_{total}$) obtained using reduced deep barcodes by min-max algorithm with VGG-16 pre-trained network. $L_1$ distance metric is used for second stage search.

| $N$ | $N_{PCA}$ values | | | |
|---|---|---|---|---|
| | 100 | 200 | 500 | 1000 |
| 20 | 60.45 | 61.66 | 60.68 | 56.68 |
| 100 | 65.06 | 65.96 | 65.13 | 62.42 |
| 200 | 66.04 | 66.79 | 66.64 | 63.40 |
| 500 | 67.09 | 66.34 | 67.09 | 66.64 |
| 1000 | 67.40 | 67.32 | 66.87 | 67.77 |

Table 6: Retrieval accuracy ($\eta_p$) obtained using reduced deep barcodes by min-max algorithm with VGG-16 pre-trained network. $L_1$ distance metric is used for second stage search.

| Method | $\eta_p$ | $\eta_w$ | $\eta_{total}$ |
|---|---|---|---|
| Deep-barcodes$^\diamond$ | 71.62 | 70.07 | 50.19 |
| Deep features$^\diamond$ | 68.91 | 67.50 | 46.51 |
| PCA$^\diamond$ | 68.53 | 67.16 | 46.03 |
| CNN* | 64.98 | 64.75 | 41.80 |
| LBP$_{(24,3)}$* | 66.11 | 62.52 | 41.33 |
| BOW* | 64.98 | 61.02 | 39.65 |

Table 6: This table compares our results with the state-of-the-art on *Kimia Path24* dataset
$^\diamond$ *Our results achieved from deep barcodes, Deep features and reduced Deep features (PCA).*
* *Best results obtained by Babaie et al. [34] using CNN (trained from scratch), LBP (24-neighbourhood, radius of 3) and BOW. Our method clearly beats state-of-the-art approach with a large margin.*